\documentclass[aps,prl,preprint,showpacs,preprintnumbers,amsmath,amssymb]{revtex4}
\usepackage{graphicx}
\graphicspath{{../}}
\usepackage{epstopdf}
\begin{document}


\title{A qualitative test for intrinsic size effect on ferroelectric phase transitions}

\author{Jin \surname{Wang}}
\email{jin.wang@epfl.ch}
\affiliation{Ceramics Laboratory, EPFL Swiss Federal Institute of Technology, CH-1015 Lausanne, Switzerland}

\author{Alexander K. \surname{Tagantsev}}
\affiliation{Ceramics Laboratory, EPFL Swiss Federal Institute of Technology, CH-1015 Lausanne, Switzerland}

\author{Nava  \surname{Setter}}
\affiliation{Ceramics Laboratory, EPFL Swiss Federal Institute of Technology, CH-1015 Lausanne, Switzerland}

\date{\today}

\begin{abstract}
The size effect in ferroelectrics is treated as a competition between the geometrical symmetry of the ferroelectric sample and its crystalline symmetry. The manifestation of this competition is shown to be polarization rotation, which is driven by temperature and/or size variations, thus providing a qualitative indication of intrinsic finite size effect on ferroelectrics. The concept is demonstrated in a simple case of PbTiO$_3$ nanowires having their axis parallel to [111]$_\mathrm{C}$ direction, where the size effect is shown to include polarization rotation and an additional phase transition. 
\end{abstract}



\maketitle
Thermodynamic phase transitions in finite systems and relevant size effects have been a focal point of interest of experimentalists and theorists for a long time. Nowadays, it is not a rarity, that an ``ordered'' (e.g. ferroelectric) state is documented and manipulated in systems 
having their smallest dimension in the range of a few lattice constants \cite{Waru,Garsia}, a scale on which the phase transition can really ``feel'' that the system is finite.
At the same time, the identification of the nature of a size effect still remains a challenge due to the multitude of factors that influence the ``ordering'' in objects with limited dimension.
For example, for ferroelectric phase transitions, one can count factors like depolarizing field and direct coupling between the polarization and the interface, which can be classified as associated with the \emph{intrinsic} size effects, and factors like size-dependent stoichiometry and microstructure, which can be classified as the \emph{extrinsic} size effects.
The experimental identification of the nature of size effects in a  system is often complicated by the fact that most of the theoretical work in the field has been focused on intrinsic scenarios, whereas experimentally, extrinsic effects often compete with those intrinsic.
On the practical level, the problem is further complicated by the fact that modeling of intrinsic size effects shows typically a lowering of the transition temperature $T_c$\cite{Tilly,Junq,Moroz,Alma}, a result that is qualitatively identical to the expected manifestation of extrinsic size effects.
Thus, to prove that the phase transition in a given system  is prone to an intrinsic size effect, one has  to analyze the functional size dependence of $T_c$, taking into account also information (typically lacking) on the impact of the extrinsic effect on $T_c$.
In this context it seems  valuable to identify  features of the intrinsic size effects that would enable their distinction on a \emph{qualitative level} from the extrinsic ones. We show below such a feature of the intrinsic size effect and illustrate it using a simple model of ferroelectric nanowires.

Consider the intrinsic size effect in a sample of a so-called ``cubic'' ferroelectric, i.e. a ferroelectric with a cubic paraelectric phase such as the classical perovskites BaTiO$_3$ and PbTiO$_3$.
In the bulk form, the ferroelectric instability in such systems can be classified as ``isotropic''.
It is controlled by the matrix of the inverse susceptibilities $\chi_{ij}$,  which is proportional to the unitary matrix:
\begin{equation}
	\label{eq:chi}
	\chi^{-1}_{ij}=\frac{\partial^2\Phi} {\partial P_i \partial P_j} =\alpha\delta_{ij}
\end{equation}
where $P_i$ stands for the Cartesian components of the polarization and $\Phi$ is the free energy density of the ferroelectric.
In the Landau theory, $\alpha$ is a temperature-dependent constant changing its sign at the point where the paraelectric phase loses its stability.
In the general case, the unstable direction with respect to appearance of the spontaneous polarization is controlled by the eigen vector of $\chi^{-1}_{ij}$ corresponding to the smallest eigen value (which, with lowering of the temperature, changes the sign first, from positive to negative).
In the case of a cubic symmetry, because of the degeneracy of the eigen values of $\chi^{-1}_{ij}$, this matrix cannot govern the directions of the ferroelectric axes of the crystal, which are controlled by higher derivatives of the free energy $\frac{\partial^4\Phi} {\partial P_i \partial P_j\partial P_k \partial P_l}$ and $\frac{\partial^6\Phi} {\partial P_i \partial P_j \partial P_k \partial P_l \partial P_m \partial P_n}$.
However the situation  changes essentially when the phase transition of a finite cubic ferroelectric sample is addressed.
The couplings that control  the intrinsic size effect can contribute to the matrix $\chi^{-1}_{ij}= \frac{\partial^2\Phi} {\partial P_i \partial P_j}$, where the free energy $\Phi$ comprises now also these couplings.
Clearly the symmetry of such a contribution is governed by that of the sample or by the superposition of the crystalline symmetry and the geometrical symmetry of the sample.
This implies that, in general, in  a finite sample, the ferroelectric instability is not isotropic any more.
It will be of the easy-plane or easy-axis type, depending on the degree of degeneracy  of the smallest eigen value of  $\chi^{-1}_{ij}$  matrix.

Here of a remarkable interest is the situation where the orientation of the symmetry elements of the $\chi^{-1}_{ij}$  matrix is different from those of the point group symmetry of the crystal.
Under such conditions, tracing  a phase transition affected by a pronounced size effect (as a function of temperature or of a certain sample dimension), one can pass from a regime where the correction to the bulk value of $\chi^{-1}_{ij}$ is large, to a regime where this correction is negligible (virtually bulk regime).
In doing so, one passes from a regime where the orientation of the polarization is controlled by the geometrical symmetry of the sample to that where it is controlled by the crystalline symmetry.
The implication is that, in general, the intrinsic size effect on phase transitions in cubic ferroelectrics should be accompanied by a rotation of the spontaneous polarization.

To illustrate this effect,  we perform a simple modeling of the ferroelectric phase transition in nanowires of PbTiO$_3$. Let us consider the ferroelectric phase transition in a free-standing  PbTiO$_3$ nanowire of a cylindrical shape with its geometrical axis parallel to the $[111]$ direction (hereafter orientations of crystallographic directions and planes will be given in the cubic reference frame of the paraelectric cubic phase).
Since, in the bulk from, the ferroelectric axes of PbTiO$_3$ are parallel to $\langle 100 \rangle$ directions, off-axis components of the polarization are expected in these nanowires.
We assume that the depolarizing field associated with these components is screened by environmental charges.
The results of experimental \cite{Wang2009,Spanier} and theoretical  \cite{Wang2009,Spanier} studies suggest that such screening can be very efficient, however it is obviously incomplete. The origin of the non-vanishing depolarizing field under screening by ambient charges is the seperation of the screening charges from the bound charges (see e.g.\cite{Gerra2006}).
The depolarizing energy associated with the incompleteness of such screening is a driving force for size effect on the phase transition of this system.
Other driving forces for this size effect are the short-range coupling of the order parameter with the interface \cite{Kretsch,Tag2008} and the surface tension\cite{Moroz}.
In order to provide a transparent illustration of the concept formulated above, we retain only the depolarizing effect in our consideration\cite{comment0} and incorparate the effect of an incomplete polarization screening in a ferroelectric nanowire within the Landau theory framework.

Addressing the problem of incomplete polarization screening in a short-circuited ferroelectric parallel-plate nano-capacitor, the depolarizing effect is described in terms of  an additional potential jump across the interface, which is proportional to the out-of-plane component of the polarization \cite{comment1} and to the effective screening length $\lambda$, which characterizes the effective thickness of the double electric layer formed by the bound and free charges \cite{Tag2008}:
\begin{equation}
	\label{eq:jump}
	\delta\phi=\frac{\lambda} {\varepsilon_0}\vec{P}\cdot\vec{n}
\end{equation}
where $\vec{n}$ is unit vector along the surface normal and $\varepsilon_0$ is the permittivity of the free space.
For the perovskite/metal interface, both simple quantum-mechanical \cite{Guro}  and advanced ab initio \cite{Junq} estimates evaluate  $\lambda$ as a fraction of Angstrom.
Assuming that for the case of screening with environmental charges  the values of $\lambda$ are comparable to those given above, we conclude that the thickness of the double electric layer formed due to the polarization screening  in  nanowires are much smaller than the typical values of their radii (2-50 nm e.g. from \cite{Spanier}).
Bearing this in mind, we apply Eq.\eqref{eq:jump} locally on the side surface of the nanowire.
Since outside the nanowire the electrostatic potential must be constant, Eq.\eqref{eq:jump} yields (to within a constant) the boundary conditions for the potential inside the nanowire.
The solution to the Laplace equation $\bigtriangledown^2\phi=0$ with such boundary conditions leads to  a uniform  depolarizing field, which is antiparallel to the polarization component $P_\perp$ normal to the  axis of the nanowire:
\begin{equation}
	\label{eq:02:m}
	\vec{E}_{\mathrm{dep}}=-\frac{\lambda}{\varepsilon_0 R}\vec{P_\perp}
\end{equation}
where $R$ is the radius of the nanowire.
The corresponding depolarizing energy reads
\begin{equation}
	\label{eq:03:m}
	\delta\Phi_\mathrm{dep}=-\frac{1}{2}\vec{E}_\mathrm{dep}\cdot\vec{P}=\frac{\lambda}{2\varepsilon_0 R}P_\perp^2
\end{equation}
The incorporation of this correction into the energy of the system modifies its instability to the easy-axis type.
In the cubic reference frame, the modified $\chi^{-1}_{ij}$ now reads
 \begin{equation}
 	\label{eq:chimod}
\chi^{-1}_{ij}=
\left(
\begin{array}{ccc}
 \alpha +\frac{2\lambda}{3\varepsilon _0R} & -\frac{\lambda}{3\varepsilon _0R} & -\frac{\lambda}{3\varepsilon _0R} \\
 -\frac{\lambda}{3\varepsilon _0R} & \alpha +\frac{2\lambda}{3\varepsilon _0R} &-\frac{\lambda}{3\varepsilon _0R} \\
 -\frac{\lambda}{3\varepsilon _0R} & -\frac{\lambda}{3\varepsilon _0R} & \alpha +\frac{2\lambda}{3\varepsilon _0R}
\end{array}
\right)		
\end{equation}
with $\alpha=\frac{T-T_0}{C \varepsilon_0}$ where $C$ and $T_0$ are the Curie-Weiss constant and Curie-Weiss temperature, respectively.
With such input, the system can be described using the following expansion for the free energy of mechanically free nanowire
\begin{widetext}
\begin{eqnarray}
\label{eq:04:energy}
	\Phi&=&\frac{1}{2}\chi^{-1}_{ij}P_iP_j+\alpha _{11}\left(P_1^4+P_2^4+P_3^4\right)+ \alpha _{12}\left(P_1^2P_2^2+P_2^2P_3^2+P_3^2P_1^2\right) +\alpha _{111}\left(P_1^6+P_2^6+P_3^6\right)\nonumber \\ &&+\;\alpha _{112}\left[P_1^4\left(P_2^2+P_3^2\right)+P_3^4\left(P_1^2+P_2^2\right)+P_2^4\left(P_1^2+P_3^2\right)\right]+\alpha _{123}P_1^2P_2^2P_3^2
\end{eqnarray}
\end{widetext}
with $\chi^{-1}_{ij}$ coming from Eq.\eqref{eq:chimod} and the high-order terms taken in the standard form for perovskite ferroelectrics.
Here, we have used the total polarization $\vec{P}$ as the order parameter in the free energy expansion, though the problems involving depolarizing effect require, strictly speaking,  the use of the ``soft-mode'' part of the polarization for this purpose \cite{Tag2008fer}.
However, one can readily show on the lines of Ref.\cite{Tag2008fer} that such substitution is legitimate if the nanowire radius $R$ is much larger than $\lambda\kappa_b$ where $\kappa_b$ is the background relative dielectric permittivity of the ferroelectric.
With a typical value of $\kappa_b$ for perovskites of about a few units \cite{Hlinka}, we expect this requirement to be  fulfilled for ferroelectric nanowires.

\begin{figure}
\begin{center}
\includegraphics[width=70mm]{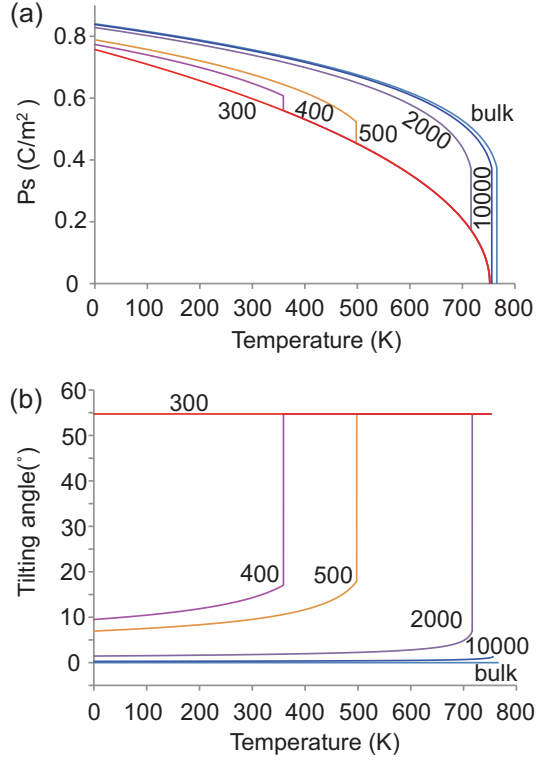}
\caption{(color online). (a) The amplitude of $P_s$ and (b) the orientation of  $P_s$ indicated by the tilting angle from [001] (tetragonal) towards [111] (rhombohedral) in the $($\={1}$10)$ plane as a function of $T$ in a PbTiO$_3$ nanowire having its geometrical axis along [111]. The accompanying numbers indicate $R/\lambda$. Note: Curves ``2000'', ``500'' and ``400'' partially overlap with the curve ``300'' at the high-temperature region; The tilting angle of $\vec{P_s}$ for $R/\lambda=300$ shows that $\vec{P_s}$ is parallel to [111].}
\label{PsT111}
\end{center}
\end{figure}

\begin{figure}
\begin{center}
\includegraphics[width=60mm]{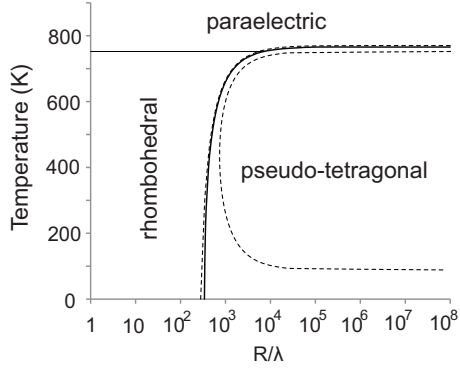}
\caption{The phase diagram of PbTiO$_3$ nanowires with the geometrical axis along [111]. The thick solid line indicates the first-order phase transition and the thin solid line corresponds to the second-order phase transition. The dashed lines are the spinodals of the first-order phase transition.}
\label{phasediagram}
\end{center}
\end{figure}

For PbTiO$_3$  wires oriented parallel to $[111]$ direction, the equilibrium states have been determined  by minimizing the free energy $\Phi$ with respect to the 3 components of the polarizations vector $\vec{P}$.
The results of the calculations  \cite{parameters} are presented in Fig.\ref{PsT111} (the absolute value of $\vec{P}$ and its orientation as functions of temperatures $T$ and the wire radius expressed in a unit of the effective screening length $R/ \lambda$) and in Fig.\ref{phasediagram} (the phase diagram of the system in $T$ - $R/ \lambda$ coordinates).
It is seen that there are three  regimes  of behavior of the system, illustrating competition between the symmetry of the geometrical sample and its crystalline symmetry.

(I) For $R/ \lambda$ smaller than some 340, it is the sample symmetry which mainly controls the system, which behaves as an easy-axis rhombohedral ferroelectric with its polarization parallel to the wire axis at all temperatures.

(II)  For $R/ \lambda$ larger than some 7200, it is  the crystalline symmetry that mainly controls the system.
It behaves as multi-axial ferroelectric with 6 possible pseudo-tetragonal orientations of the polarization, which is very close to those in bulk PbTiO$_3$.
In this regime the size effect manifests itself in a weak rotation of the spontaneous polarization in $\{110\}$ planes passing through [111] direction and a shift of the transition temperature $T_{\mathrm{C}}$ from its bulk value down to $T_0$.

(III) For intermediate values of $R/ \lambda$, a crossover regime takes place. Here, a first-order transition (driven by variations of temperature or/and wire radius) between the rhombohedral and  pseudo-tetragonal phases takes place.
In this case, a jump-like rotation of the polarization at the transition and a further substantial rotation in the pseudo-tetragonal phase (driven by variations of $T$ and $R/ \lambda$) occur as a manifestation of the competition between the geometrical and crystalline symmetries.
The spinodals of this transition are  also shown in Fig.\ref{phasediagram}

\begin{figure}
\begin{center}
\includegraphics[width=55mm]{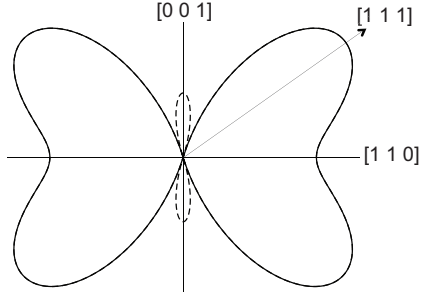}
\caption{The magnitude of the $P^4$-contribution to the free energy for a fixed value of $P$ as a function of the orientation of $\vec{P}$ in the $\{$\={1}$10\}$ plane in PbTiO$_3$. The dashed and solid curves indicate negative and positive sign respectively.}
\label{2Dplotbeta}
\end{center}
\end{figure}

A remarkable feature of the phase diagram shown in Fig.\ref{PsT111}a is that the ferroelectric phase transition between the paraelectric and the rhombohedral  phases is of a second order whereas in the bulk material, the transition is of a first order.
The origin of this feature can be elucidated using the polar plot (Fig.\ref{2Dplotbeta}) of the magnitude of the $P^4$-contribution to the free energy for a fixed value of $P$ as a function of the orientation of $\vec{P}$.
It is known that the sign of this magnitude controls the order of the transition.
It is seen that, for $\langle 100 \rangle$ directions, the sign is negative, controlling the first-order phase transition in bulk PbTiO$_3$.
At the same time, for $\langle 111 \rangle$ directions, the sign is positive, explaining the second-order phase transition in ``thin'' nanowires where the depolarizing effect urges the appearance of polarization parallel or antiparallel to  $[111]$ direction.

It is interesting to estimate, based on Fig.\ref{PsT111}, the dimensions of nanowires that are likely to manifest a strong size effect. If one assumes that the screening length is 0.1 $\mathrm{\AA}$, the curves representing $R/\lambda$ of 300, 400, 500, and 10000 represent wire diameters of 6, 8, 10, and 200 nm, respectively. In other words, a nanowire of a realistic 8 nm diameter is predicted to show, upon cooling from the paraelectric phase, a second-order phase transition to a rhombohedral ferroelectric phase, followed by another phase transition (rhombohedral to monoclinic) at about 80$\,^\circ \mathrm{C}$ (Fig.\ref{PsT111}). Experimental results are not available at present for PbTiO$_3$ nanowires with growth direction along $\langle 111 \rangle$. In the case of Pb(Zr$_{0.15}$Ti$_{0.85}$)O$_3$, monodomain wires were obtained, having  [001] parallel to the nanowire axis \cite{Jin}. However, albeit having a rather large diameter (d$\approx$50 nm), polydomain nanowires of KNbO$_3$ \cite{Dkhil}, having $\langle 100 \rangle$ direction along the wire axis, showed a low temperature monoclinic phase instead of the rhombohedral phase that occurs in bulk KNbO$_3$, which might be interpreted as a fingerprint of an intrinsic size effect in line of the arguments outlined above.  

All in all, the presented model clearly illustrates the qualitative feature of the intrinsic size effect on phase transitions in finite samples, i.e. an order parameter rotation occurs in the cases where the geometry of the sample lowers the symmetry of the system compared to the crystalline symmetry.
Despite the simplicity of the model we believe that it provides a qualitatively correct picture of the phase transition in  PbTiO$_3$ nanowires of $[111]$ orientation \cite{comment2}.
Keeping in mind a fraction of Angstrom as an estimate for $\lambda$, the model predicts, for realistic dimensions ($R$ lying in the range of several to few tens of nm), essential rotation of the spontaneous polarization caused by variations of the temperature and wire radius $R$. It is important to note that short-range polarization/interface coupling and elastic effects related to the surface tension could also break the isotropic nature of the ferroelectric instability, similarly leading to a rotation of the spontaneous polarization during the phase transition in cases where these effects manifest themselves.
As the rotation of the spontaneous polarization entails a rotation of the spontaneous deformation, this qualitative manifestation of the size effect might be detected by XRD analysis.

\begin{acknowledgments}
This project was supported by the Swiss National Science Foundation.
\end{acknowledgments}

\newpage

\begin{thebibliography}{999}
\bibitem{Waru}  M. P. Warusawithana, C. Cen, C. R. Sleasman, J. C. Woicik, Y. L. Li, L. F. Kourkoutis, J. A. Klug, H. Li, P. Ryan, L. P. Wang, M. Bedzyk, D. A. Muller, L. Q. Chen, J. Levy, and D. G. Schlom, Science, \textbf{324}, 367 (2009).
\bibitem{Garsia} V. Garcia, S. Fusil, K. Bouzehouane, S. Enouz-Vedrenne, N. D. Mathur, A. Barthelemy, and M. Bibes, Nature, \textbf{460}, 81 (2009).
\bibitem{Tilly} D.R. Tilley and B. Zeks, Solid State Communications \textbf{49}, 823 (1984).
\bibitem{Junq} J. Junquera and P. Ghosez, Nature, \textbf{422}, 506 (2003).
\bibitem{Moroz} A. N. Morozovska, E. A. Eliseev, and M. D. Glinchuk, Physical Review B, \textbf{73}, 214106 (2006).
\bibitem{Alma} E. Almahmoud, I. Kornev, and L. Bellaiche, Physical Review B, \textbf{81}, 064105 (2010).
\bibitem{Wang2009} R. V. Wang, D. D. Fong, F. Jiang, M. J. Highland, P. H. Fuoss, C. Thompson, A. M. Kolpak, J. A. Eastman, S. K. Streiffer, A. M. Rappe, and G. B. Stephenson, Physical Review Letters, \textbf{102}, 047601 (2009).
\bibitem{Spanier} J. E. Spanier, A. M. Kolpak, J. J. Urban, I. Grinberg, O. Y. Lian, W. S. Yun, A. M. Rappe, and H. Park, Nano Letters, \textbf{6}, 735 (2006).
\bibitem{Kretsch} R. Kretschmer and K. Binder, Phys. Rev. B, \textbf{20}, 1065 (1979).
\bibitem{Tag2008} A.K. Tagantsev, G. Gerra, and N. Setter, Physical Review B, \textbf{77}, 174111 (2008).
\bibitem{comment0}
 Addressing the situation with the  incomplete screening, customarily one does not call upon the effect of the short-range polarization/inteface coupling (see e.g.\cite{Stengel2009}).
Recent first principles calculations justifies such a neglect of the short-range polarization/inteface coupling at least in  some systems \cite{Tag2008}. Second, the strength of the surface tension is basically unknown because of the missing information on its magnitude of the perovskites whereas to get appreciable effects one has to assume quite high values of the latter.
\bibitem{Stengel2009} M. Stengel, D. Vanderbilt, and N. A. Spaldin, Nature Materials, \textbf{8}, 392 (2009).
\bibitem{Gerra2006} G. Gerra, A.K. Tagantsev, N. Setter, and K. Parlinski, Physical Review Letters, \textbf{96}, 107603 (2006).
\bibitem{comment1}
Strictly speaking, the additional potential jump across the interface is controlled by the normal component of electrical displacement $D$.
Here, following Ref. \cite{Tag2008} we neglect the difference between $P$ and $D$. 
\bibitem{Guro} G. M. Guro, I. I. Ivanchik, and N. F. Kovtonyu, Soviet Physics Solid State,USSR, \textbf{11}, 1574 (1970).
\bibitem{Tag2008fer} A.K. Tagantsev, Ferroelectrics, \textbf{375}, 19 (2008)
\bibitem{Hlinka} J. Hlinka and P. Marton, Physical Review B, \textbf{74}, 104104 (2006).
\bibitem{parameters} The parameters (in SI units, the temperature $T$ in $^\circ \mathrm{C}$ ) used in the calculation: $\alpha=7.6 (T-479)\times10^5$, $\alpha_{11}=-7.3\times10^7$, $\alpha_{12}=7.5\times 10^8$, $\alpha_{111}=2.6\times 10^8$, $\alpha_{112}=6.1\times10^8$, $\alpha_{123}=-3.7\times 10^9$ (taken from N. A. Pertsev, A. G. Zembilgotov, and A. K. Tagantsev, Physical Review Letters, \textbf{80}, 1988 (1998)).
\bibitem{Jin} J. Wang, C. S. Sandu, E. Colla, Y. Wang, W. Ma, R. Gysel, H. J. Trodahl, N. Setter and M. Kuball, Applied Physics Letters, \textbf{90}, 133107 (2007)
\bibitem{Dkhil} L. Louis, P. Gemeiner, I. Ponomareva, L. Bellaiche, G. Geneste, W. Ma, N. Setter and B. Dkhil, Nano Letters, \textbf{10}, 1177 (2010)
\bibitem{comment2}
The model is simplified also in the sense that  the polydomain pseudo-tetragonal state is not taken into account as competing with the depolarizing-effect-affected single-domain pseudo-tetragonal state.
As was shown for an electroded system, the polydomain state can be more energetically favorable \cite{Brat} than single-domain affected by the depolarizing effect.
If this happens, the first-order transition line should be replaced with that between the rhombohedral and polydomain pseudo-tetragonal state.
Of importance is that the size-effect-driven rotation of the polarization is expected in this case as well.
\bibitem{Brat} A. M. Bratkovsky and A. P. Levanyuk, Journal of Computational and Theoretical Nanoscience, \textbf{6}, 465 (2009).
\end{thebibliography}
\end{document}